\documentclass[]{aastex631}
\usepackage{amsmath}

\hypersetup{linkcolor=blue,citecolor=blue,filecolor=cyan,urlcolor=blue}

\submitjournal{ApJ}
\accepted{18 November 2021}

\definecolor{DCW}{rgb}{1,0.6,0}

\begin{document}

\title{Examining Two-Dimensional Luminosity-time Correlations for Gamma Ray Burst Radio Afterglows with VLA and ALMA}

\correspondingauthor{Maria Dainotti}
\email{maria.dainotti@nao.ac.jp}

\author[0000-0003-3411-6370]{Delina Levine}
\affiliation{Department of Astronomy, University of Maryland, College Park, MD 20742, USA}

\author[0000-0003-4442-8546]{Maria Dainotti}
\affiliation{National Astronomical Observatory of Japan, 2-21-1 Osawa, Mitaka, Tokyo 181-8588, Japan}
\affiliation{Space Science Institute, Boulder, CO, USA}
\affiliation{The Graduate University for Advanced Studies, SOKENDAI, Shonankokusaimura, Hayama, Miura District, Kanagawa 240-0193, Japan}
\footnote{First and second authors contributed equally to this paper}

\author[0000-0002-1786-6252]{Kevin J. Zvonarek}
\affiliation{Department of Physics, University of Michigan, Ann Arbor, MI 48109, USA}

\author[0000-0002-0173-6453]{Nissim Fraija} 
\affiliation{Instituto de Astronomía, Universidad Nacional Autónoma de México Circuito Exterior, C.U., A. Postal 70-264, 04510 México D.F., México}

\author[0000-0002-3222-9059]{Donald C. Warren}
\affiliation{RIKEN Interdisciplinary Theoretical and Mathematical Sciences Program (iTHEMS), Wak\={o}, Saitama, 351-0198 Japan}

\author[0000-0002-0844-6563]{Poonam Chandra}
\affiliation{National Centre for Radio Astrophysics, Tata Institute of Fundamental Research, Ghaneshkhind Pune 411007, India}

\author[0000-0003-1707-7998]{Nicole Lloyd-Ronning}
\affiliation{Computational Physics and Methods Group (CCS-2), Los Alamos National Lab, Los Alamos, NM, USA 87545}
\affiliation{Department of Science and Engineering, University of New Mexico, Los Alamos, 87544}

\begin{abstract}
Gamma-ray burst (GRB) afterglow emission can be observed from sub-TeV to radio wavelengths, though only 6.6\% of observed GRBs present radio afterglows. We examine GRB radio light curves (LCs) to look for the presence of radio plateaus, resembling the plateaus observed in X-ray and optical. We analyze 404 GRBs from the literature with observed radio afterglow and fit 82 GRBs with at least 5 data points with a broken power law (BPL) model, requiring 4 parameters. From these, we find 18 GRBs that present a break feature resembling a plateau. We conduct the first multi-wavelength study of the Dainotti correlation between the luminosity $L_a$ and the rest-frame time of break $T_a^*$ for those 18 GRBs, concluding that the correlation exists and resembles the corresponding correlation in X-ray and optical wavelengths after correction for evolutionary effects. We compare the $T_a^*$ for the radio sample with $T_a^*$ values in X-ray and optical data \citep{2013ApJ...774..157D, 2020ApJ...905L..26D}, finding significantly later break times in radio. We propose that this late break time and compatibility in slope suggests either a long-lasting plateau or the passage of a spectral break in the radio band. We also correct the distribution of the isotropic energy $E_{\rm iso}$ vs. the rest-frame burst duration $T^*_{90}$ for evolutionary effects and conclude that there is no significant difference between the $T^*_{90}$ distribution for the radio LCs with a break and those without.
\end{abstract}

\section{Introduction} \label{sec:intro}

Gamma-ray bursts (GRBs) are observed in all wavelengths, from high energy gamma-rays to radio. GRBs are characterized by an energetic ``prompt emission'' of $\gamma$-rays followed by a much longer period of lower-energy emission called the ``afterglow'', lasting from hundreds of seconds to years. X-ray, optical, and radio afterglows are not observed equally: X-ray afterglows have been detected in $\sim 66\%$ of observed GRBs when we consider the full sample of detected GRBs including the GRBs which are not only observed by the Neihl Gehrels Swift Observatory (Swift), optical afterglows in 38\%, and radio afterglows in only 6.6\% of all known GRBs (Greiner 2021).\footnote{\url{https://www.mpe.mpg.de/~jcg/grbgen.html}} Indeed, some GRBs are too faint to be detected in radio \citep{2012ApJ...746..156C}. With the new Square Kilometer Array (SKA) facilities \citep{2021ExA....51....1B} and SKA pathfinder \citep{2007PASA...24..174J, 2011apmc.conf.1178S} we will be able to observe the radio afterglows of more GRBs. Out of the total number of GRBs observed in radio, we count that the majority (152) are observed by the Very Large Array (VLA).

Swift light curves (LCs) resulting from GRB afterglows have highlighted complicated features inconsistent with a simple power law decay \citep{2007ApJ...669.1115S, 2009ApJ...703.1696Z}. Analysis of X-ray LCs has shown the existence of plateaus, or a flattening in the afterglow emission between the prompt emission and the subsequent afterglow decay \citep{2007ApJ...669.1115S, 2013ApJ...774..157D, 2020arXiv200311252F, 2020ApJ...905..112F}. These plateaus have also been confirmed in optical LCs \citep{2020ApJ...905L..26D}.

An interesting two-dimensional correlation between the luminosity, $L_a$, and the rest-frame end time of the plateau, $T_a^*$, known as the Dainotti correlation, was discovered more than a decade ago \citep{2008MNRAS.391L..79D, 2011ApJ...730..135D, 2013ApJ...774..157D, 2015MNRAS.451.3898D, 2016ApJ...825L..20D, 2017A&A...600A..98D, 2017ApJ...848...88D, 2020ApJ...904...97D, 2020ApJ...905L..26D, 2020ApJ...903...18S} and has been proposed as a tool to standardize the plateau sample of GRBs. These plateaus are thought to be produced by continuous energy injection. One proposed explanation is accretion falling back onto a black hole, where energy is released into the external shock, interacting with the surrounding medium and injecting energy into the observed afterglow \citep{2007ApJ...670..565L, 2012MNRAS.426L..86O}. Another interpretation involves the spin-down luminosity from a newborn magnetar providing the continuous energy injection \citep{1993ApJ...408..194T, 1992ApJ...392L...9D, 1992Natur.357..472U, 2001ApJ...552L..35Z, 2011MNRAS.413.2031M, 2014MNRAS.443.1779R, 2015ApJ...813...92R, 2018ApJ...869..155S, 2021ApJ...907...78F}.

Using the magnetar scenario, \citet{2013ApJ...774..157D} and \citet{2020ApJ...905L..26D} showed that this relation can be seen in both X-ray and optical wavelengths with a slope of $\approx -1$. This relation in X-ray has been used to build a GRB Hubble diagram with redshift values up to $z > 8$ \citep{2009MNRAS.400..775C, 2010MNRAS.408.1181C, 2014ApJ...783..126P, 2013ApJ...774..157D}. If this correlation also exists in radio, it could reveal information about the underlying GRB emission mechanisms and prove a step toward the standardization of the varied GRB population.

Our goal is to determine the existence of radio plateaus and examine the two-dimensional Dainotti correlation in radio. To our knowledge, this is the first such analysis with radio data, and thus the most complete multi-wavelength study of this relation. Our paper is organized as follows: in section~\ref{sec:sample}, we describe our data sample. In section~\ref{sec:results}, we discuss the multi-wavelength Dainotti correlation, distribution of the isotropic energy in the prompt emission, $E_{\rm iso}$ and the rest frame time duration of the prompt emission $T^*_{90}$, and correlation between $E_{\rm iso}$ vs. the rest frame end time of the plateau emission, $T_a^*$ and its correspondent luminisosity, $L_a$, as well as a comparison of $T_a^*$ in X-ray, optical, and radio. In section~\ref{sec:discussion}, we discuss the implications of the results achieved and present our conclusions.

\section{Data Selection} \label{sec:sample}

We take our sample from all published radio afterglows in the literature, mainly observed by the Very Large Array (VLA). The largest portion of our data comes from \citet{2012ApJ...746..156C}, consisting of 304 radio afterglows observed from 1997 to 2011. We extend our search to 2020, gathering an additional 100 GRBs from the literature for a total sample of 404 GRBs. We also note that of this sample, four GRBs have been observed by NAOJ-affiliated telescopes - two by the Atacama Large Millimeter Array (ALMA), one by the Nobeyama 45-m Radio Telescope, and one by the East Asian VLBI Network (EAVN).

We then put our sample through a filtering process to obtain LCs useful for our analysis. We first reject all radio observations reporting only upper limits, bringing our sample to 211 GRBs. To attempt a fit to the radio LCs, we require at least five observations within the same frequency, discarding 127 GRBs that do not fit that criteria. We further discard 2 GRBs without known redshift, leaving us with a final sample of 82 GRBs. 

Where the data is available, we consider multiple LCs within different frequency bands for one GRB. We therefore attempt to fit 202 LCs in total from our sample of 82 GRBs - on average, we fit LCs in 2-3 frequencies for each GRB, though some have more LCs (i.e. GRB030329, has 9 LCs.) We perform the fit with a broken power law (BPL) model, using the equation: 
\begin{equation}
    F(t)  =
    \begin{cases}
    F_a (\frac{t}{T_a})^{-\alpha_1} & t < T_a \\
    F_a (\frac{t}{T_a})^{-\alpha_2} & t \geq T_a\,,
    \end{cases}
\end{equation}
where $F_a$ is the flux at the end of the plateau emission in $\mathrm{erg}$ $\mathrm{cm}^{-2}$ $\mathrm{s}^{-1}$, $T_a$ is the observed frame time in seconds at the end of the plateau emission, and $\alpha_1$ and $\alpha_2$ refer to the temporal power law decay indices before and after the break, respectively. The $\alpha_1$ index is important because it determines the flatness of the plateau. We compute the flux in these units for ease of comparison to X-ray and optical data. 

To compile our final sample, we first discard 75 LCs because the data of the LCs are too scattered to be fitted with a BPL. Then, we discard 18 LCs, because they do not support the shape of the BPL with a plateau (i.e. their $|\alpha_1|>0.5$, or they can be fitted with a simple power law, etc.). 
In addition, 70 LCs don't fulfill the \citet{1978A&A....66..307A} prescription regarding the $\Delta \chi^2$ analysis, namely, we varied the $\Delta \chi^2$ so that the 1 $\sigma$ bounds could be determined assuming that the $\Delta \chi^2$ shape is a parabola for each of the parameters involved in the fitting, see \citet{1978A&A....66..307A} for additional details.

\begin{figure*}[t!]
\gridline{\fig{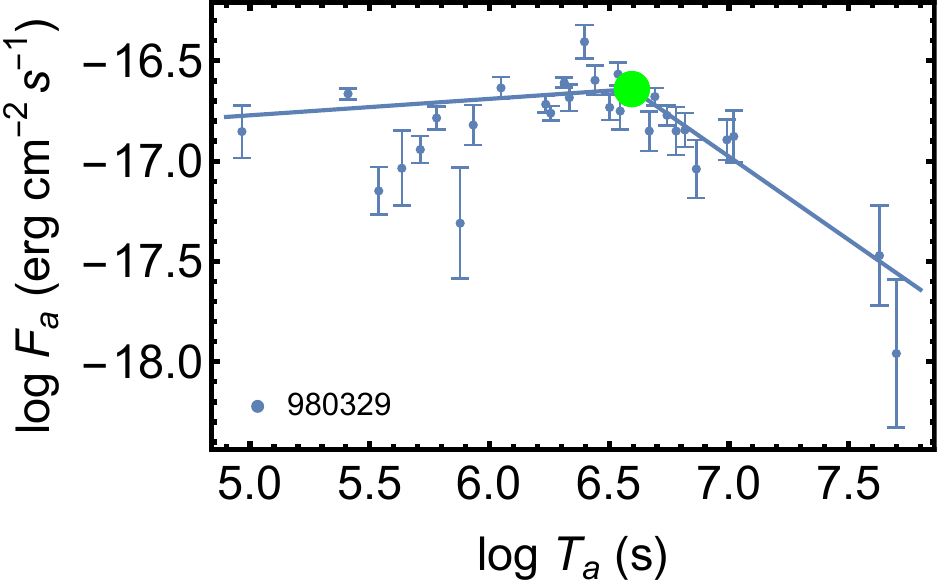}{0.28\textwidth}{}
\fig{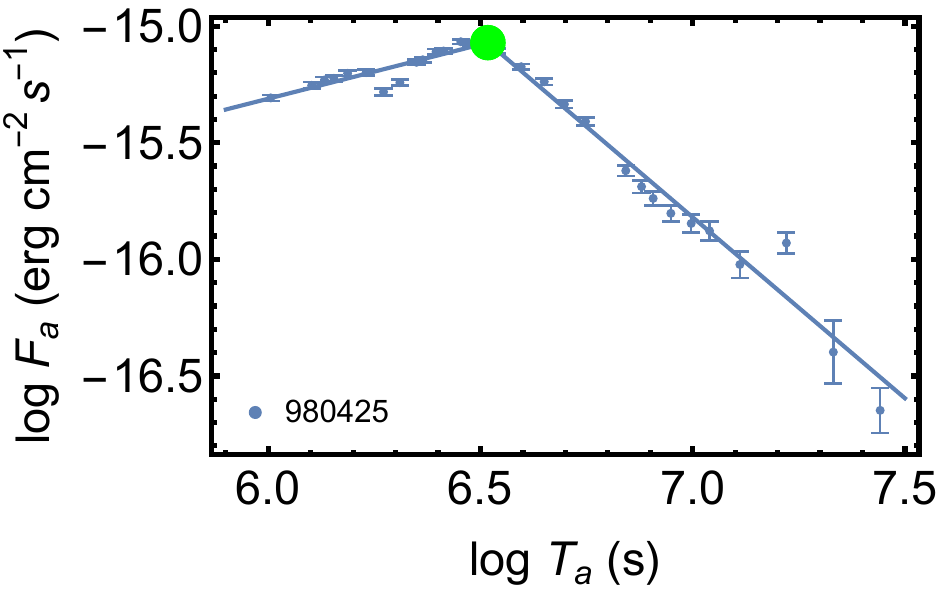}{0.28\textwidth}{}
\fig{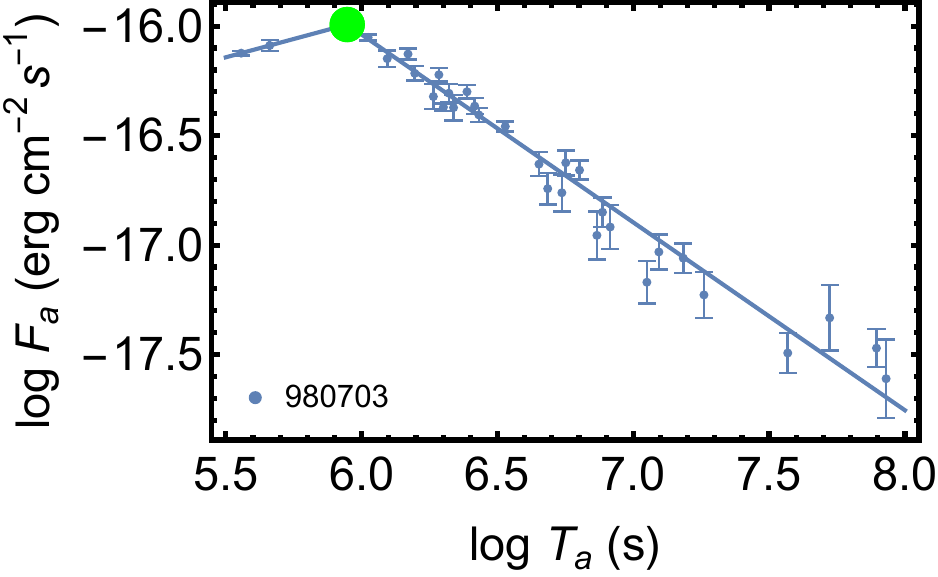}{0.28\textwidth}{}}
\gridline{\fig{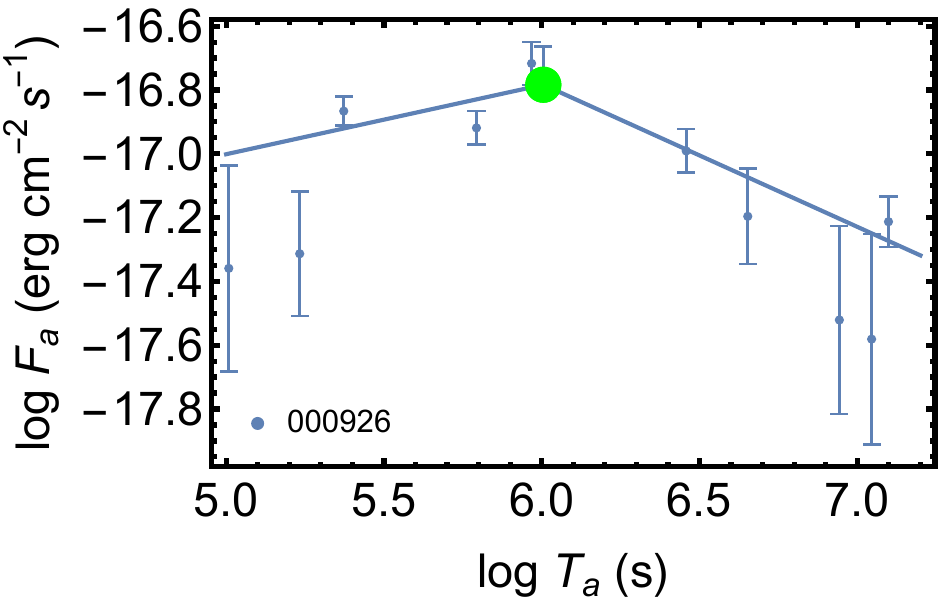}{0.28\textwidth}{}
\fig{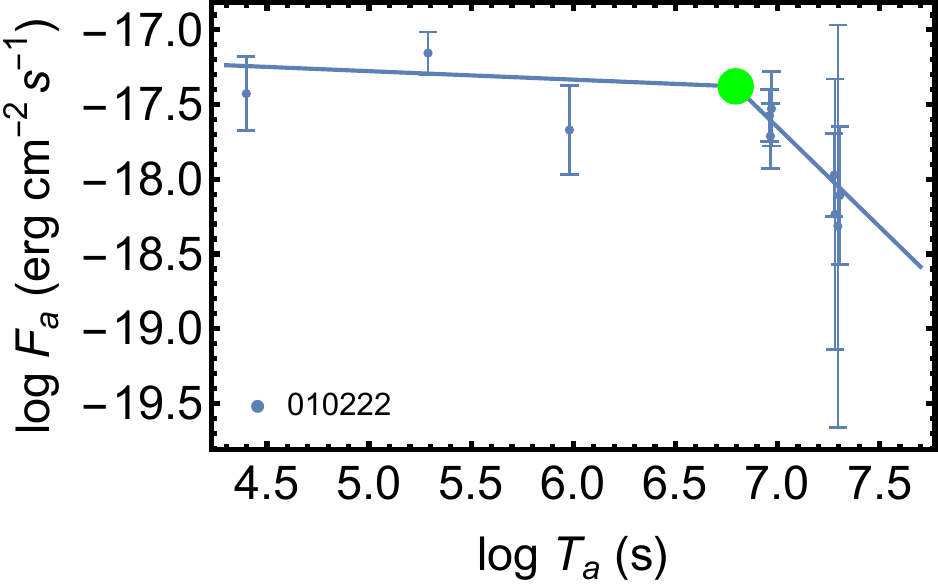}{0.28\textwidth}{}
\fig{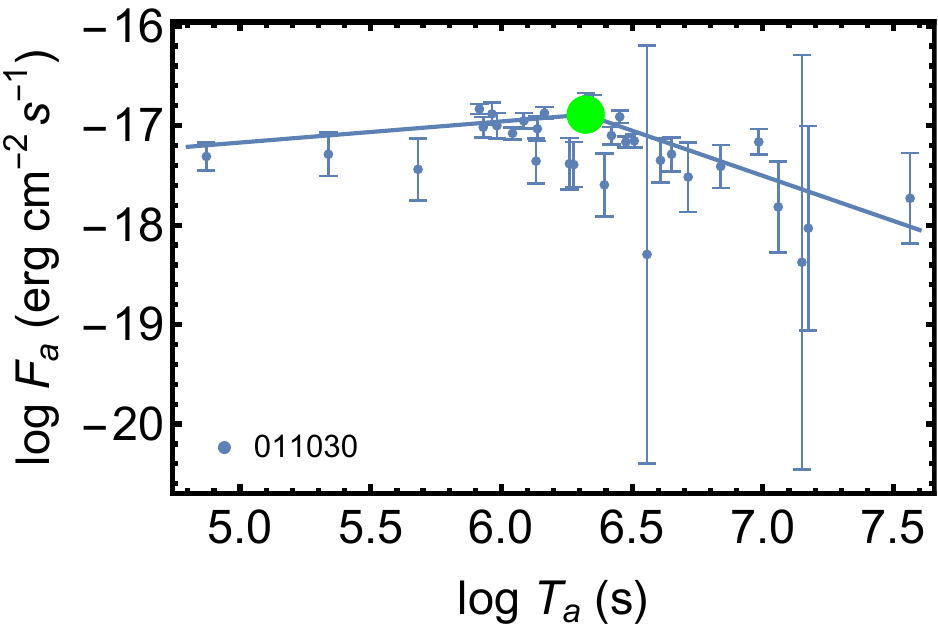}{0.28\textwidth}{}}
\gridline{\fig{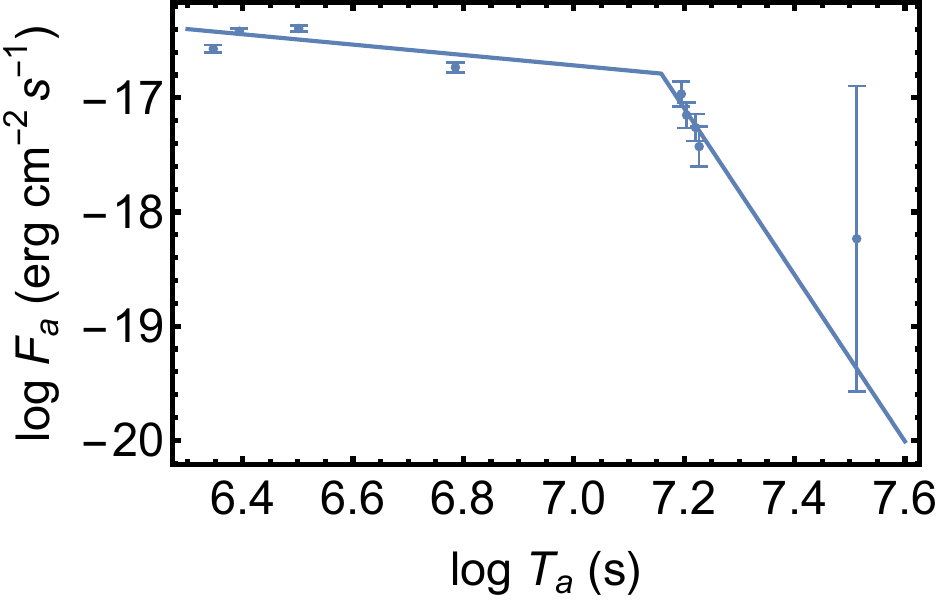}{0.28\textwidth}{}
\fig{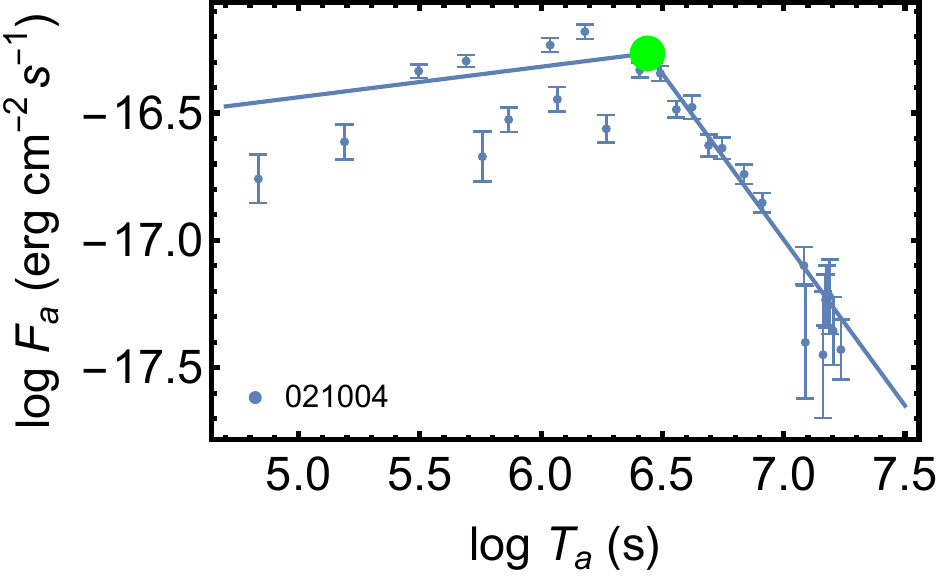}{0.28\textwidth}{}
\fig{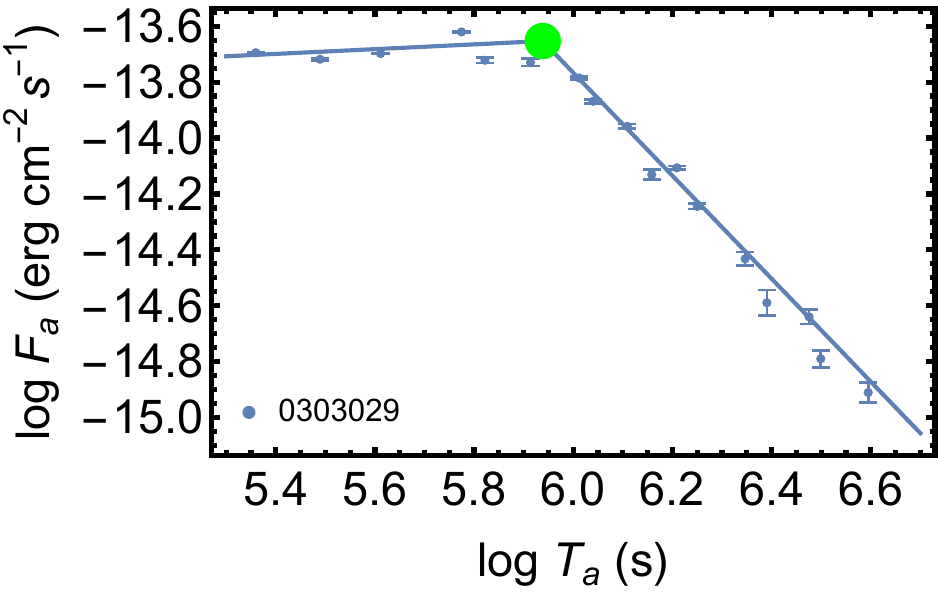}{0.28\textwidth}{}}
\gridline{\fig{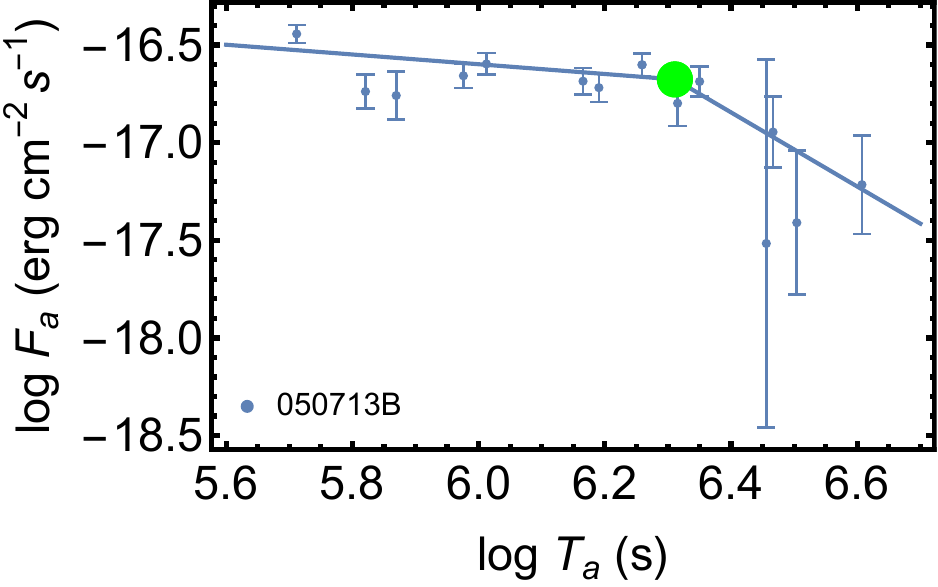}{0.28\textwidth}{}
\fig{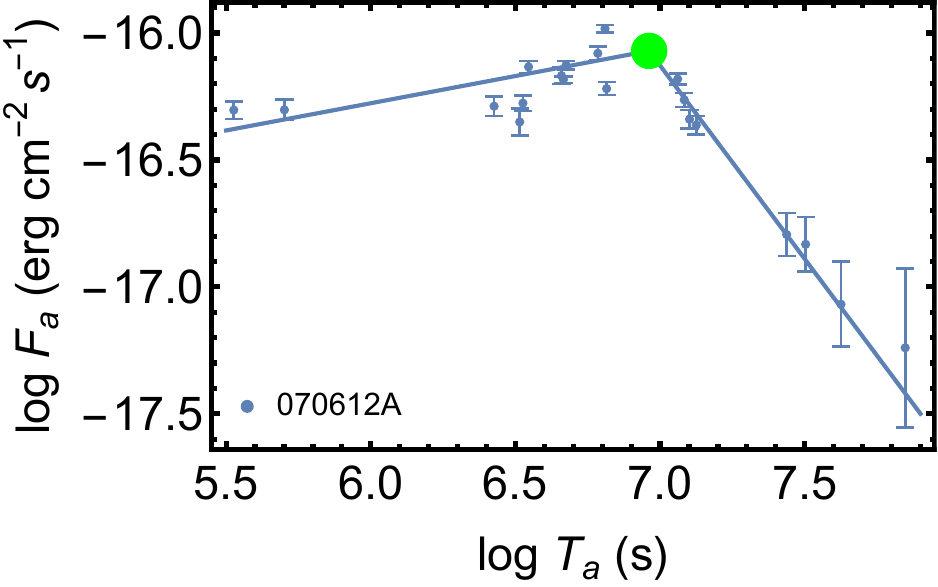}{0.28\textwidth}{}
\fig{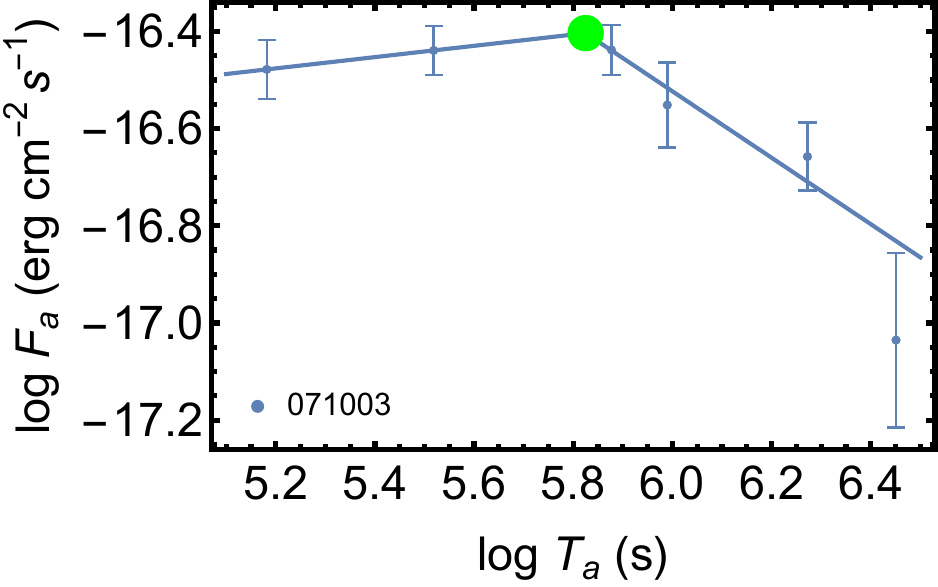}{0.28\textwidth}{}}
\gridline{\fig{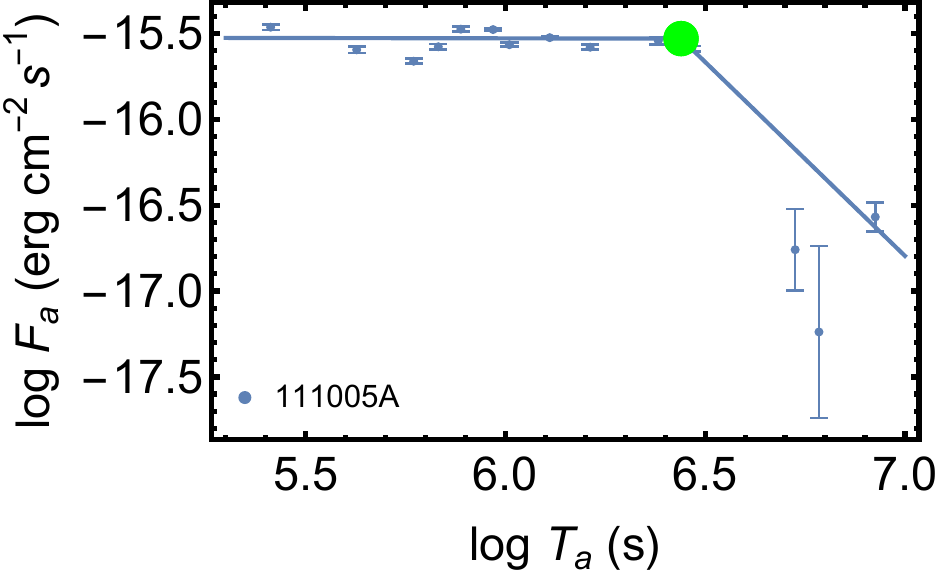}{0.28\textwidth}{}
\fig{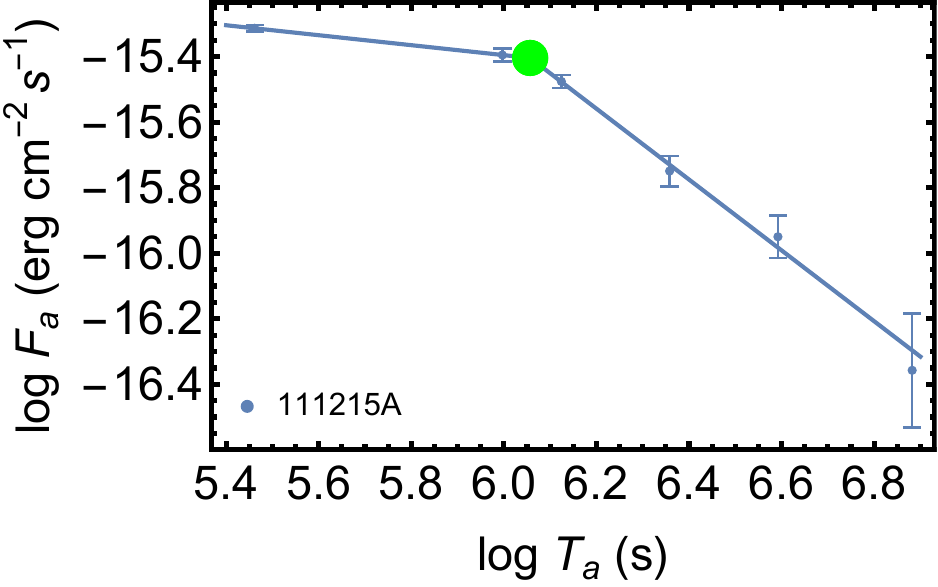}{0.28\textwidth}{}
\fig{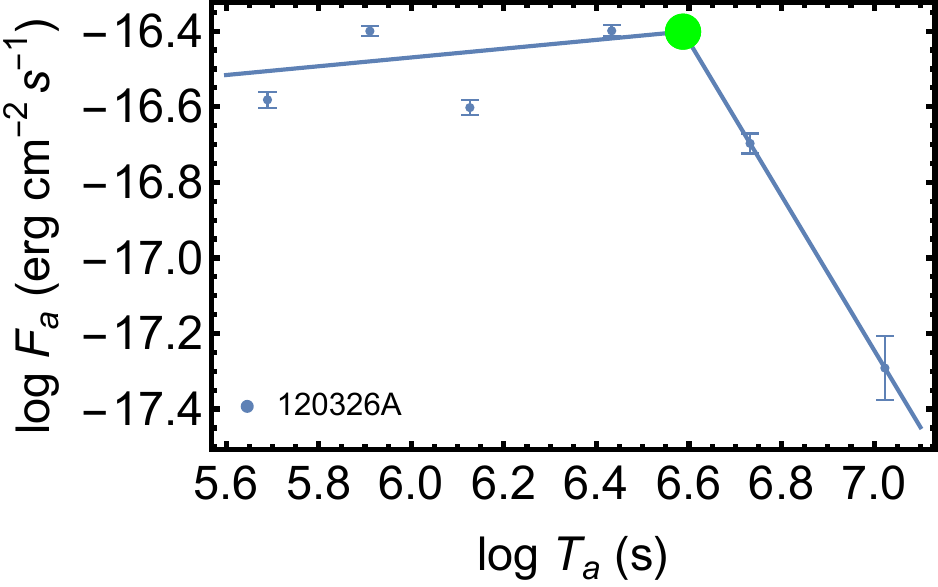}{0.28\textwidth}{}}
\gridline{\fig{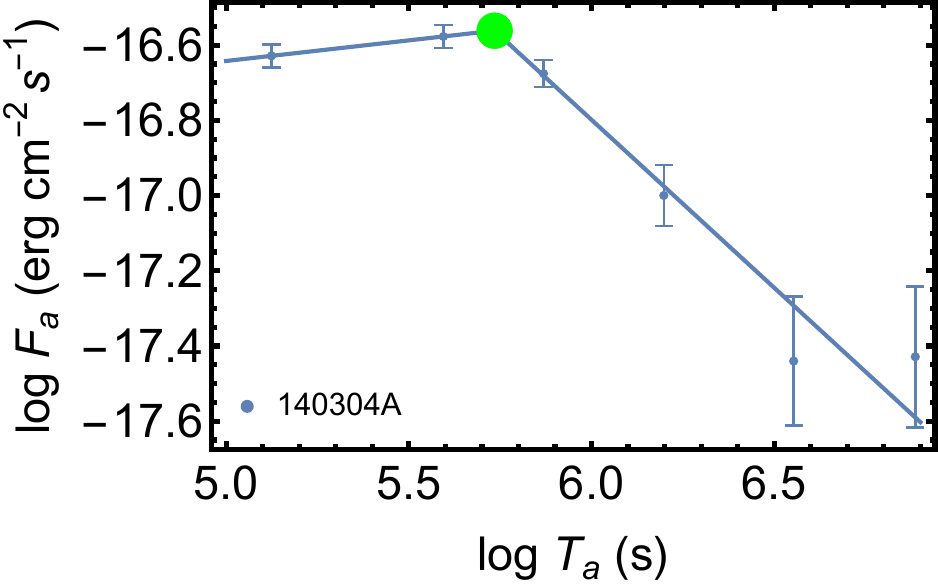}{0.28\textwidth}{}
\fig{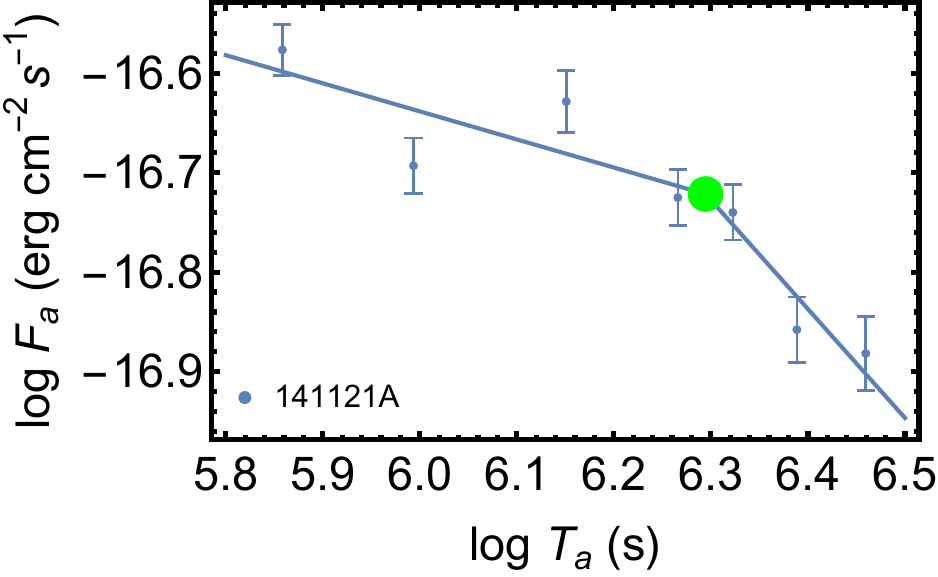}{0.28\textwidth}{}
\fig{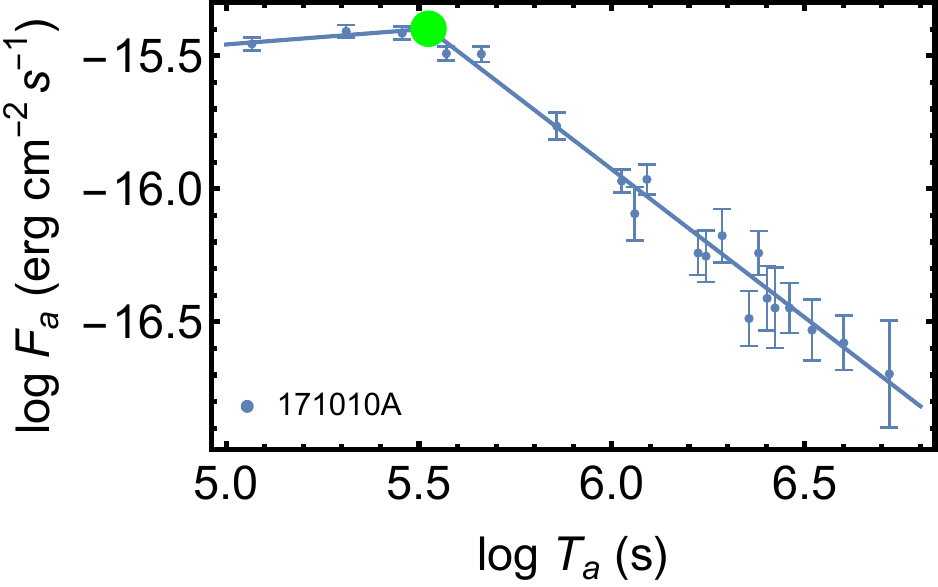}{0.28\textwidth}{}}
\caption{LCs accepted to plateau sample from BPL fitting.\label{fig:LCs1}}
\end{figure*}

With this selection we are left with 39 LCs that display a clear break - however, we further reject 9 too steep to present a plateau, with $|\alpha_1|> 0.5$, and 12 from repeated GRBs. Thus, we find 18 LCs that resemble a plateau with $0<|\alpha_1|<0.5$. We report the best-fit parameters for the sample of 18 plateau GRBs in Table~\ref{tab:table1}. All the errors quoted in this paper and gathered in Table 1 are calculated to 1 $\sigma$. We show the LCs of the plateau sample in figure \ref{fig:LCs1}.

Of these 18 GRBs, all are classified as long GRBs, with (GRB 020903) classified as an X-ray flash (XRF), 1 (GRB 141121A) considered X-ray rich (XRR), and 2 with Supernova associations \citep{2017A&A...600A..98D}. We classify GRBs with supernova associations according to the convention presented in \citet{2012grb..book..169H} in relation to the GRB-SNe Ic connection, with GRB030329 and GRB 980425 classified as ``type-A" or SN-A, indicating strong spectroscopic evidence for the association.

\begin{figure*}[ht!]
\gridline{\fig{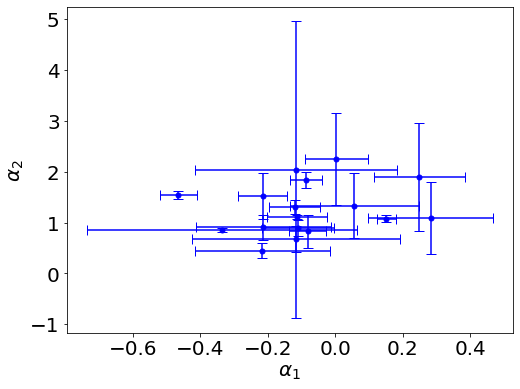}{0.5\textwidth}{}
\fig{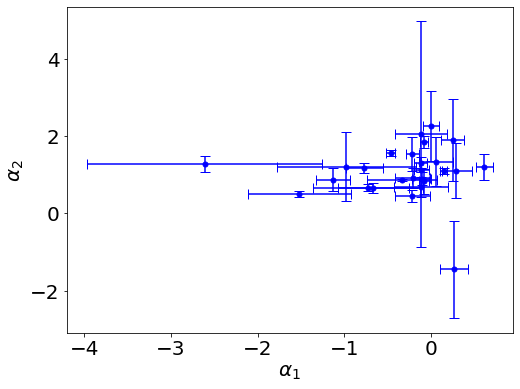}{0.5\textwidth}{}}
\caption{Left: $\alpha_2$ vs. $\alpha_1$ for ``plateau" sample. Right: distribution for ``plateau" sample plus the ``break" sample. \label{fig:alphas}}
\end{figure*}

We note that two GRBs, GRB 020903 and GRB 120326A, have very large error bars in the $\alpha_2$ parameter - thus, we do not consider them in the subsequent analysis of the Dainotti correlation. However, we still include them in the plateau sample, as the sample size is small for the computation of the Efron \& Petrosian method. In addition,those GRBs meet all other criteria for acceptance of the fit. A plot of $\alpha_2$ vs. $\alpha_1$ for the sample of 16 GRBs with a plateau used in the correlation analysis is shown in the left panel of figure \ref{fig:alphas}. The same relation for the plateau sample plus the 9 additional GRBs with $|\alpha_1| > 0.5$, hereafter referred to as the ``break" sample (as they can be reliably be fitted with a broken power law model, but the slope is too steep to be considered a plateau), is shown in the right panel of Fig. \ref{fig:alphas}.

From the observed radio flux, we then compute the luminosity $L_{a}$ at the time of break, $T_a$, using the equation: 
\begin{equation}
L_{a}=4\pi D_L^2 (z) F_{a} (T_a)K\,,
\end{equation}
where $F_{a}$ is the observed flux at $T_a$, $D_L^2 (z)$ is the luminosity distance assuming a flat $\Lambda \textrm{CDM}$ model with $\Omega_M=0.3$ and $H_0=70$ $\mathrm{km}$ $\mathrm{ s^{-1}}$ $\mathrm{Mpc^{-1}}$, and K is the k-correction: 
\begin{equation}
K=\frac{1}{(1+z)^{\alpha_1-\beta}}\,,
\end{equation}
with $\beta$ as the radio spectral index of the GRB \citep{2012ApJ...746..156C}. The $\beta$ values were gathered from the literature; where no value existed, the average of the existing values, $\beta = 0.902$, was assigned, with the average of the known uncertainties $\sigma_\beta = 0.17$.

\section{Results} \label{sec:results}
\subsection{Luminosity-time correlation} \label{sec:LT}

\begin{figure*}[ht!]
\epsscale{1.15}
\gridline{\fig{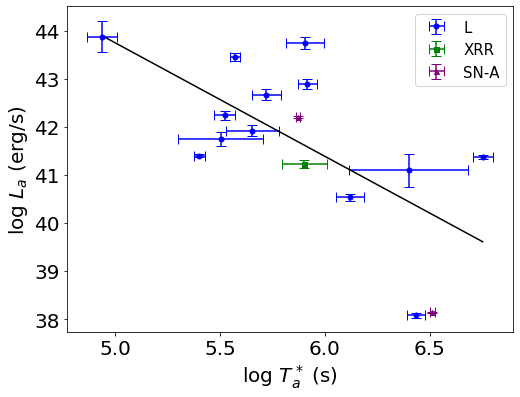}{0.5\textwidth}{}
\fig{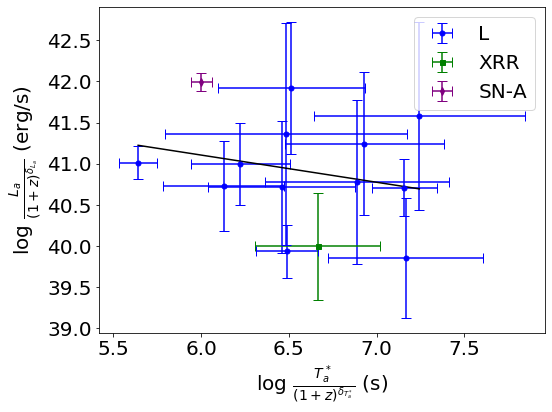}{0.5\textwidth}{}}
\gridline{\fig{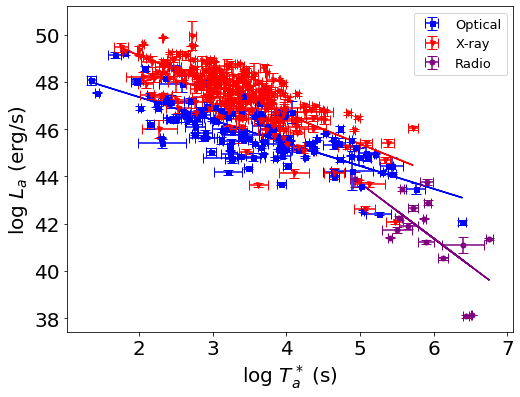}{0.5\textwidth}{}
\fig{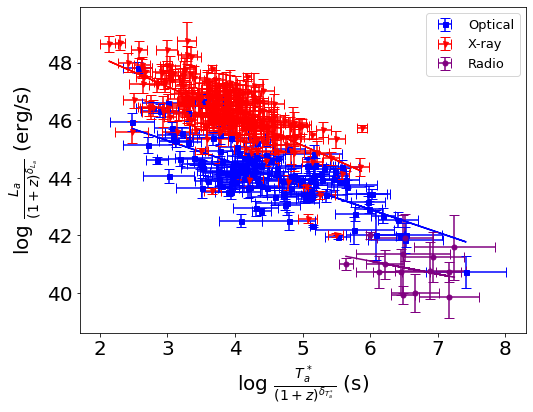}{0.5\textwidth}{}}
\caption{Top left: uncorrected $L_a-T_a^*$ correlation for 16 LCs, with GRBs differentiated by class: ``L" refers to long GRBs, ``XRR" corresponds to X-ray rich GRBs, and ``SN-C" and ``SN-A" correspond to type-A and type-C supernovae as described in section \ref{sec:sample}. Upper right: correlation corrected for selection biases, and with low-redshift outliers removed. Lower left: multi-wavelength $L_a -T_a^*$ correlation in X-ray, optical and radio shown in blue, red, and purple respectively. Lower right panel: correlation corrected for evolutionary effects with the same color code. 
\label{fig:correlations}}
\end{figure*}

For the LCs resembling a plateau, we examine the Dainotti correlation between $L_a$ and the rest-frame time of break $T_a^{*} = \frac{T_a}{(1+z)}$ (the star * denotes the rest frame), similar to the work in \citet{2017ApJ...848...88D} and \citet{2020ApJ...905L..26D} (figure~\ref{fig:correlations}). For a review on the afterglow correlations see \citet{2017NewAR..77...23D, 2018PASP..130e1001D, 2018AdAst2018E...1D}. We use the Bayesian D'Agostini method with the {\tt\string cobaya} Python package to obtain our fitting parameters. Uncertainties are given to 1 $\sigma$. The luminosity-time (Dainotti) correlation in radio is defined as:
\begin{equation}
\textrm{log} L_{a,radio}=C_o+a_{rad}\times \textrm{log} T_a^*\,,
\end{equation}
where $C_o$ is the normalization constant and $a_{\rm rad}$ is the slope determined by the linear fit. Using the sub-sample of 16 GRBs from the full plateau sample, we find best-fit parameters $C_o= 55.42 \pm 3.91 $ and $a_{\rm rad} = -2.34 \pm 0.66$. An ANOVA test gives a p-value that this correlation is drawn by chance of $p = 0.005$, and the Spearman $\rho$ coefficient for this correlation is $\rho = -0.6$, indicating that the correlation is significant. 

We compare the results of the $L_a-T_a^*$ correlation in radio to the corresponding correlations in X-ray and optical \citep{2013ApJ...774..157D, 2020ApJ...905L..26D}. We take our sample of 222 X-ray LCs from \citet{2013ApJ...774..157D}, Dainotti et al. (2021a in preparation) and our sample of 131 optical LCs from \citet{2020ApJ...905L..26D}, Dainotti et al. (2021b in preparation). To our knowledge, this is the first time such a comparison has been considered. We find a slope in X-ray of $a_X = -1.25 \pm 0.07$ and in optical of $a_{opt} = -0.97 \pm 0.07$. The radio slope agrees with these values within 2.1 $\sigma$. 

However, for a comparison that accounts for biases and redshift evolution, we apply the Efron-Petrosian (EP) method \citep{1992ApJ...399..345E} to recover their intrinsic slopes of the full sample of 18 plateau GRBs due to the the paucity of the data \citep{2013ApJ...774..157D, 2015ApJ...800...31D, 2015MNRAS.451.3898D}. To this end, we need to mimic the evolution of the variables with redshift with a simple function of redshift, $f(z)=(1+z)^\delta$, where $\delta$ is the slope of the evolutionary function determined by the EP method through the computation of a modified version of the Kendall $\tau$ statistics. The slope $\delta$ is found when $\tau = 0$, corresponding to the removal of the evolution. We update the analysis of the evolution in the X-ray sample (Dainotti et al. 2021a in preparation) and we find $T_a^{*'} = \frac{T_a^*}{(1+z)^{\delta_{T_a^*}}}$ where $\delta_{T_a^*} = -1.2 \pm 0.28$ and $ L_a^{'} = \frac{L_a}{(1+z)^{\delta_{L_a}}}$ where $\delta_{L_a} = 2.4 \pm 0.65$. For an updated optical sample (Dainotti et al. 2021b in preparation), we find $\delta_{T_a^*} = -2.1 \pm 0.60$, which agrees with the X-ray value within 1.5 $\sigma$, and $\delta_{L_a} = 3.97 \pm 0.45$, which agrees within 2.4 $\sigma$. This is more likely due to a difference in sample size than a result of an underlying physical process.

For the radio data, the luminosity limit {has been determined} using a method described in \citet{2021ApJ...914L..40D}, in which a complete ``parent'' sample of GRBs with known peak fluxes are compared to a sub-sample of GRBs with known peak flux and known redshift. Here, the peak radio flux is defined as the highest flux observed in the LC, which coincides with $T_a^*$ for the majority of the ``break" sample. We therefore take the same convention for the parent sample. 

The two-sample Kolmogorov-Smirnov (KS) test is then used to quantify the probability, as a function of flux limit, that the sub-sample is pulled from the parent sample. We find $>90\%$ probability for all flux limits, with an observed increase in p-value beginning at $f_{lim} = -17.8$, in units of erg~cm$^{-2}$~s$^{-1}$, and reaching a plateau at 100\% probability at $f_{lim} = -17.2$. Therefore, we choose -17.2 as the radio luminosity limit and find $\delta_{T_a^*} = -1.94 \pm 0.86$, which agrees with the value in X-ray and optical within 1 $\sigma$, and $\delta_{L_a} = 3.15 \pm 1.65$, which also agrees within 1 $\sigma$.

After applying the EP method using this prescription for limiting luminosity, we find the slope for the Dainotti correlation in radio as $a_{\rm rad} = -0.26 \pm 0.71$. This is compatible with the corresponding correlation in X-ray, with a corrected slope of $a_{\rm X}= -1.02 \pm 0.07$, within 1.07 $\sigma$, and the corrected slope in optical, $a_{\rm opt}= -0.79 \pm 0.06$, within 0.75 $\sigma$. However, we note that two GRBs, GRB 980425 and GRB 111005A, appear to be outliers, with lower radio luminosity and redshift than the rest of the sample - if those GRBs are removed from the radio correlation, the slope of the corrected correlation becomes $a_{\rm rad} = -0.45 \pm 0.47$. This value of the correlation, which we consider as the intrinsic value, has been corrected for the effects of selection bias and redshift evolution and does not consider systematically different, low-luminosity GRBs. This slope agrees with the slope of the corrected correlation in X-ray within 1.26 $\sigma$ and the corrected optical correlation within 0.72 $\sigma$.
We here stress that these two outliers are the closest GRBs at the smallest redshift - an order of magnitude less than the redshift of GRB 030329. It is highly likely that these two GRBs are off-axis GRBs and this is the reason why their luminosities are substantially lower than the ones expected at their specified redshift, for additional details see \citet{2015ApJ...799....3R}. The radio Dainotti correlation for the plateau sample is shown without the EP method corrections in the upper left panel of figure \ref{fig:correlations}, and with the correction (and removal of outliers) in the upper right panel. The multi-wavelength correlations in X-rays, optical and radio, without corrections (left) and with corrections and removal of outliers (right), are shown in the lower panels. The X-ray, optical and radio data are shown in red, blue, and  purple, respectively.

Looking at the distribution of radio data within the luminosity-time correlation by class (figure~\ref{fig:correlations}), we observe no particular clustering of any type of GRBs. 

\subsection{$E_{\rm iso} - T^*_{90}$ distribution and the presence of LC breaks} \label{sec:ET}

\begin{figure*}[ht!]
\gridline{\fig{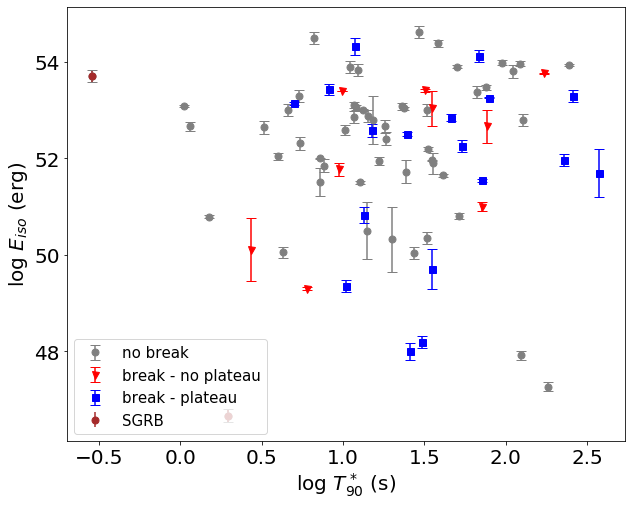}{0.5\textwidth}{}
\fig{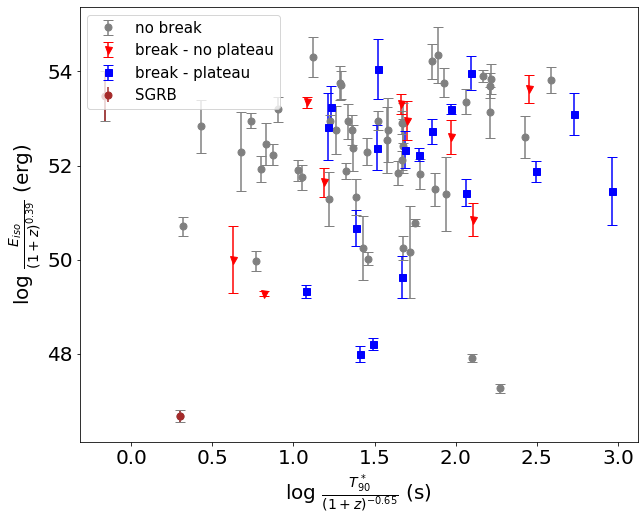}{0.5\textwidth}{}}
\gridline{\fig{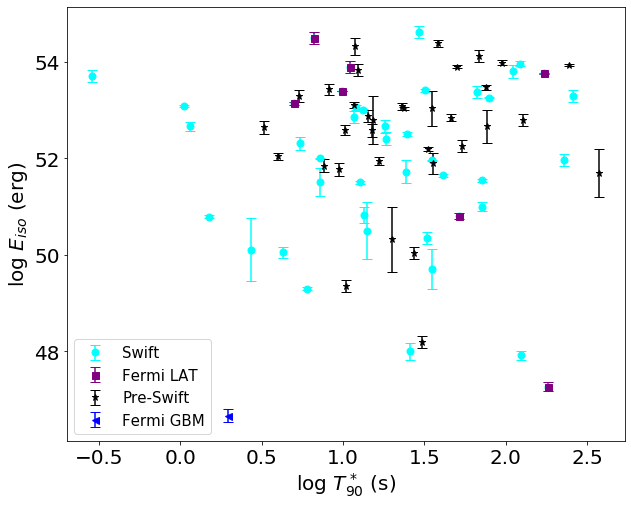}{0.5\textwidth}{}
\fig{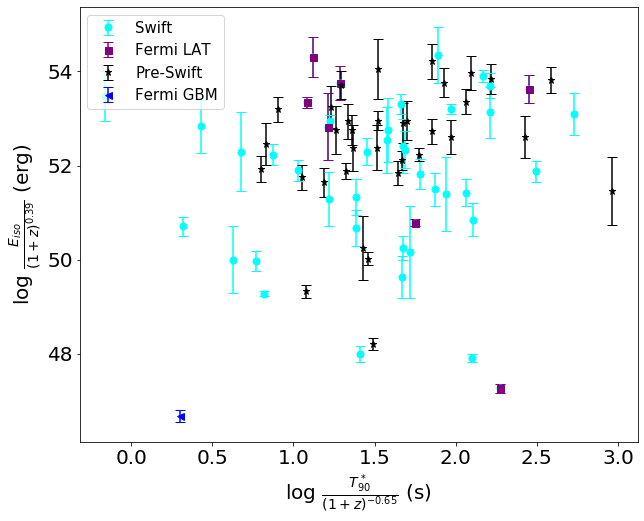}{0.5\textwidth}{}}
\caption{Left panel: $E_{\rm iso}-T^{*}_{90}$ distribution color-coded according to whether the sample includes the break. Right panel: distribution color-coded by instrument. The lower left and right panel show the corrected distributions correspondent to the above panels. 
\label{fig:ET90}}
\end{figure*}

To further investigate the behavior of the GRBs that present a break, we examine whether there is a relation between the existence of the break and the energy and duration of the prompt emission properties of a GRB. Thus, we investigate the distribution of the isotropic energy, $E_{\rm iso}$ vs. the rest-frame burst duration, $T^*_{90} = \frac{T_{90}}{(1+z)}$ for a subsample of 80 GRBs of the 82 GRBs considered for fitting, for which we could either find a value of $E_{\rm iso}$ in the literature or compute the value from the literature - where no value could be found (GRBS 050509C and 170105A), we compute the $E_{\rm iso}$ using the following equation:
\begin{equation}
E_{\rm iso}=4\pi D_L^2 (z) S K\,,
\end{equation}
where $S$ is the fluence, $D_L^2 (z)$ is defined as in equation (2), and $K$ is the correction 
\begin{equation}
K=\frac{1}{(1+z)^{1-\beta}}\,,
\end{equation}
with $\beta$ as the spectral index of the GRB. The $S$ and $\beta$ values are taken from the Swift/BAT GRB catalog \footnote{\url{https://swift.gsfc.nasa.gov/results/batgrbcat/index_tables.html}}. We plot the distribution of the $E_{\rm iso}$ vs. $T^*_{90}$, shown in Figure~\ref{fig:ET90}. In the upper panels, we color-code the sample according to whether it presents a break – the blue refers to the 18 GRBs that present a break resembling a plateau, with $0<|\alpha_1|<0.5$; the red refers to the 9 GRBs that present a break, but with steeper $|\alpha_1|>0.5$, and the grey refers to 53 GRBs that do not present a break. In the lower panels, we note from which satellite and instruments the GRBs have been observed: the Fermi Large Area Telescope (LAT), the Fermi Gamma Ray Burst Monitor (GBM), and Swift/BAT are shown in purple, blue, and cyan, respectively. The left panels show the GRB variables without the EP method correction, the right panels show the variables after correction for selection bias and redshift evolution.

Similar to the analysis performed for the Dainotti relation in radio shown in the previous section, we correct the $E_{iso} - T^*_{90}$ distribution for selection biases and redshift evolution using the EP method, which gives $E_{\rm iso}^{'}=\frac{E_{\rm iso}}{(1+z)^{\delta_{E_{\rm iso}}}}$ and $T_{90}^{*'}=\frac{T^*_{90}}{(1+z)^{\delta_{T^*_{90}}}}$ where $\delta_{E_{\rm iso}} = 0.39 \pm 0.88$ and $\delta_{T^*_{90}} = -0.65 \pm 0.27$. The $\delta_{T^*_{90}}$ value agrees with values previously reported in \citet{2019MNRAS.488.5823L, 2020MNRAS.494.4371L} within 1 $\sigma$, while the $\delta_{E_{\rm iso}}$ agrees within 2.17 $\sigma$. We show the corrected distribution in the lower two panels of figure~\ref{fig:ET90}, and find no particular trend or clustering after correction. An examination of the corrected $T^*_{90}$ distribution alone (figure \ref{fig:Tdist}) shows overlap between the GRBs with a break and those without. A KS test between the two samples yields $KS=0.18$ with $p=0.56$, suggesting that they are drawn from the same parent sample.

\subsection{Correlation of $E_{\rm iso}$ vs. $L_a$ and $T_{a}^*$}
\begin{figure*}[ht!]
\gridline{\fig{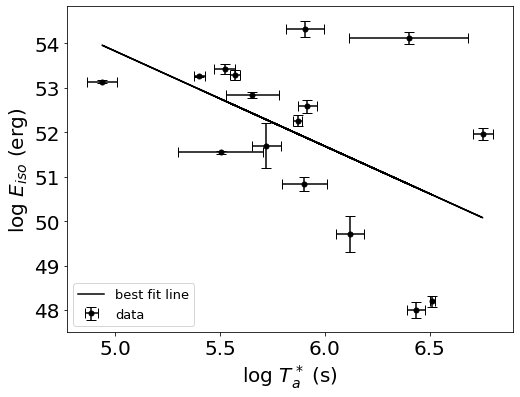}{0.5\textwidth}{}
\fig{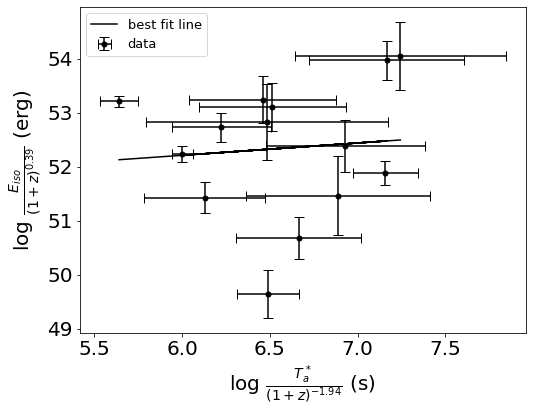}{0.5\textwidth}{}}
\gridline{\fig{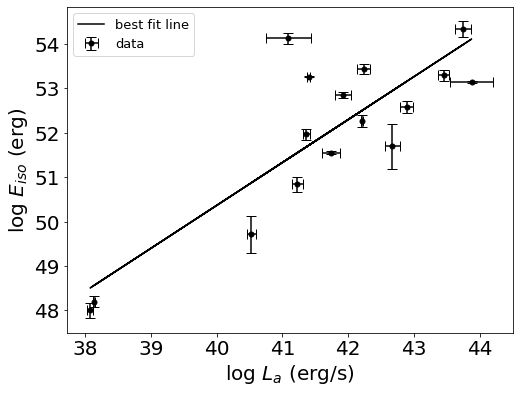}{0.5\textwidth}{}
\fig{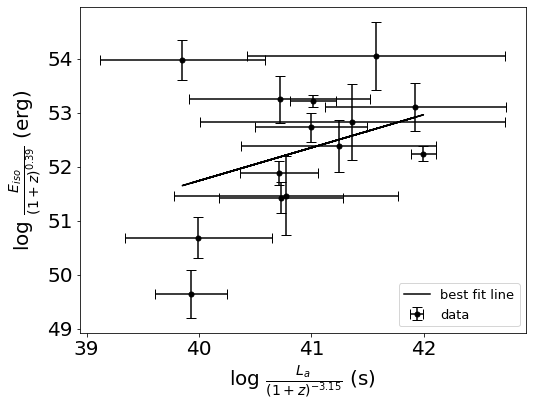}{0.5\textwidth}{}}
\caption{Top left: Uncorrected $E_{\rm iso}$ vs. $T_a^*$ for sample of 16 GRBs with a plateau considered in section \ref{sec:LT}. Top right:  $E_{\rm iso}$ vs. $T_a^*$ after correction for evolutionary effects. Bottom panels: Uncorrected and corrected $E_{\rm iso}$ vs. $L_a$ correlation for sample of 14 GRBs considered in the corrected Dainotti correlation, shown in left and right panels, respectively.
\label{fig:ETL}}
\end{figure*}

To better understand the relation of the prompt emission to the radio afterglow, we further analyze the correlation of $E_{\rm iso}$ vs. $L_a$ and $E_{iso}$ vs. $T_a^*$, similar to the work done in \citet{2011MNRAS.418.2202D} in X-rays. For the sample of 16 GRBs with a plateau considered for the $L_a$-$T_a^*$ correlation without correction, we find that the slope of the correlation between $E_{\rm iso}$ and $T_a^*$ is $-2.14 \pm 0.96$, while the slope of the correlation between $E_{\rm iso}$ and $T_a^*$ is $0.97 \pm 0.17$. After correction for evolutionary effects and removal of the two low-luminosity GRBs, we find the slope after correction for $E_{\rm iso}$ vs. $T_a^*$ is $0.23 \pm 0.7$, while the slope for the corrected $E_{\rm iso}$ vs. $L_a$ correlation is $0.61 \pm 0.42$. This indicates that both correlations are susceptible to evolutionary effects - indeed, the correlation between $E_{\rm iso}$ vs. $T_a^*$ nearly vanishes after correction, indicating that the original result is very likely a result of selection bias and redshift evolution. These results are shown in fig \ref{fig:ETL}, with $E_{\rm iso}$ vs. $T_a^*$ in the top panels and $E_{\rm iso}$ vs. $L_a$ in the lower panels. The left panels show the uncorrected correlation, while the corrected correlation is shown in the right panels.

\subsection{Comparison of $T_a^*$ in X-ray, optical and radio}
\begin{figure*}[ht!]
\gridline{\fig{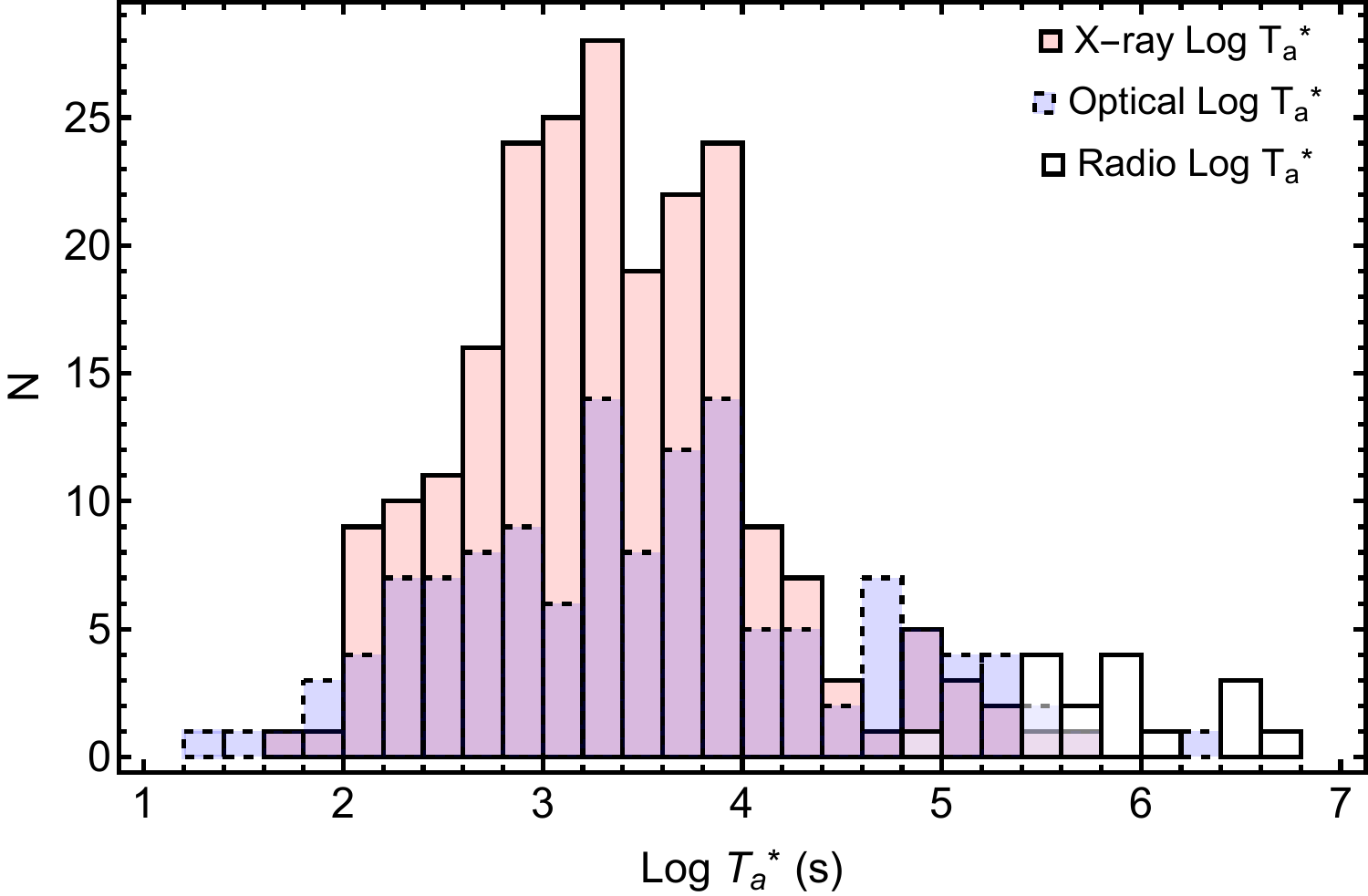}{0.5\textwidth}{}
\fig{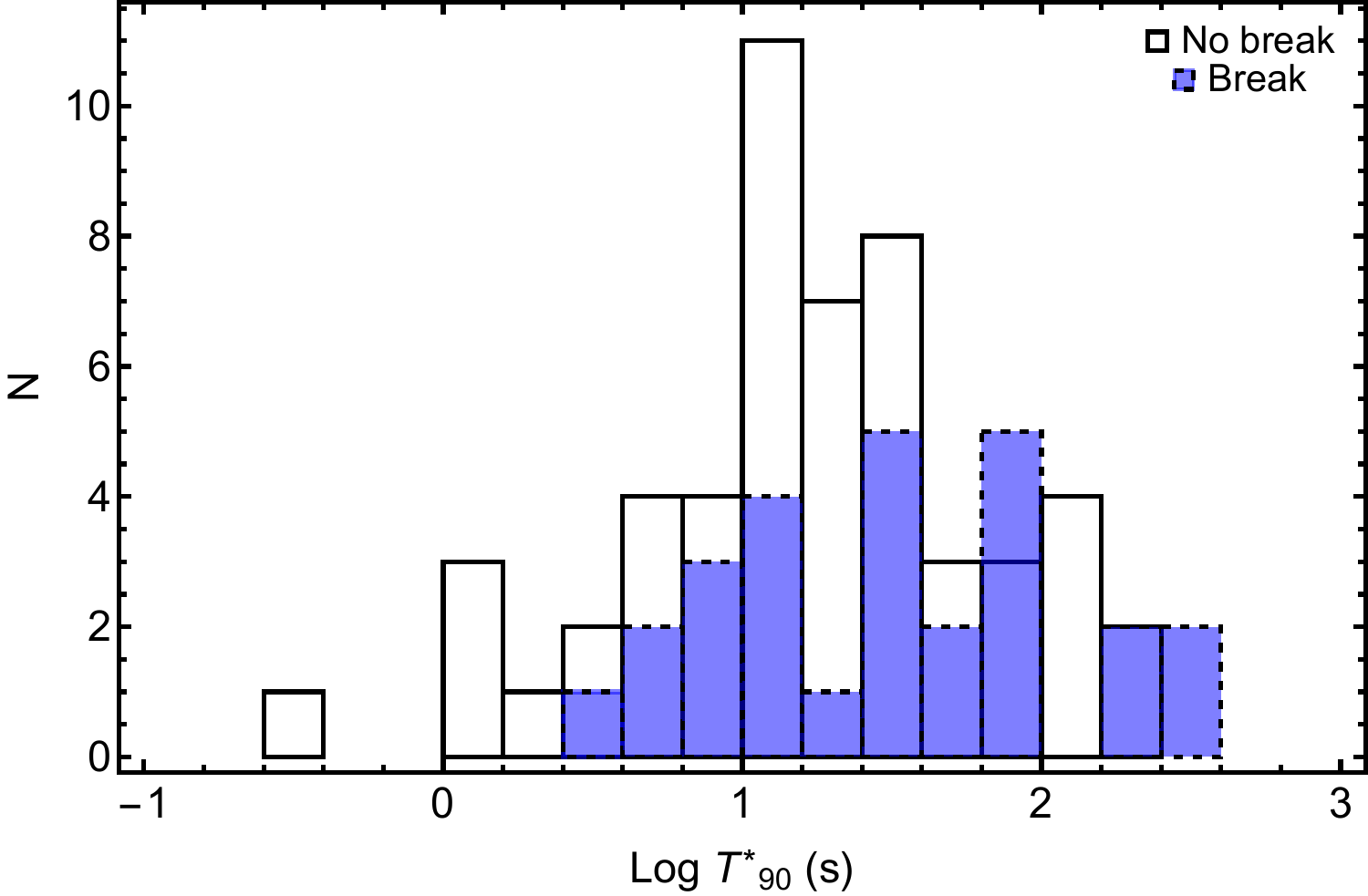}{0.5\textwidth}{}}
\gridline{\fig{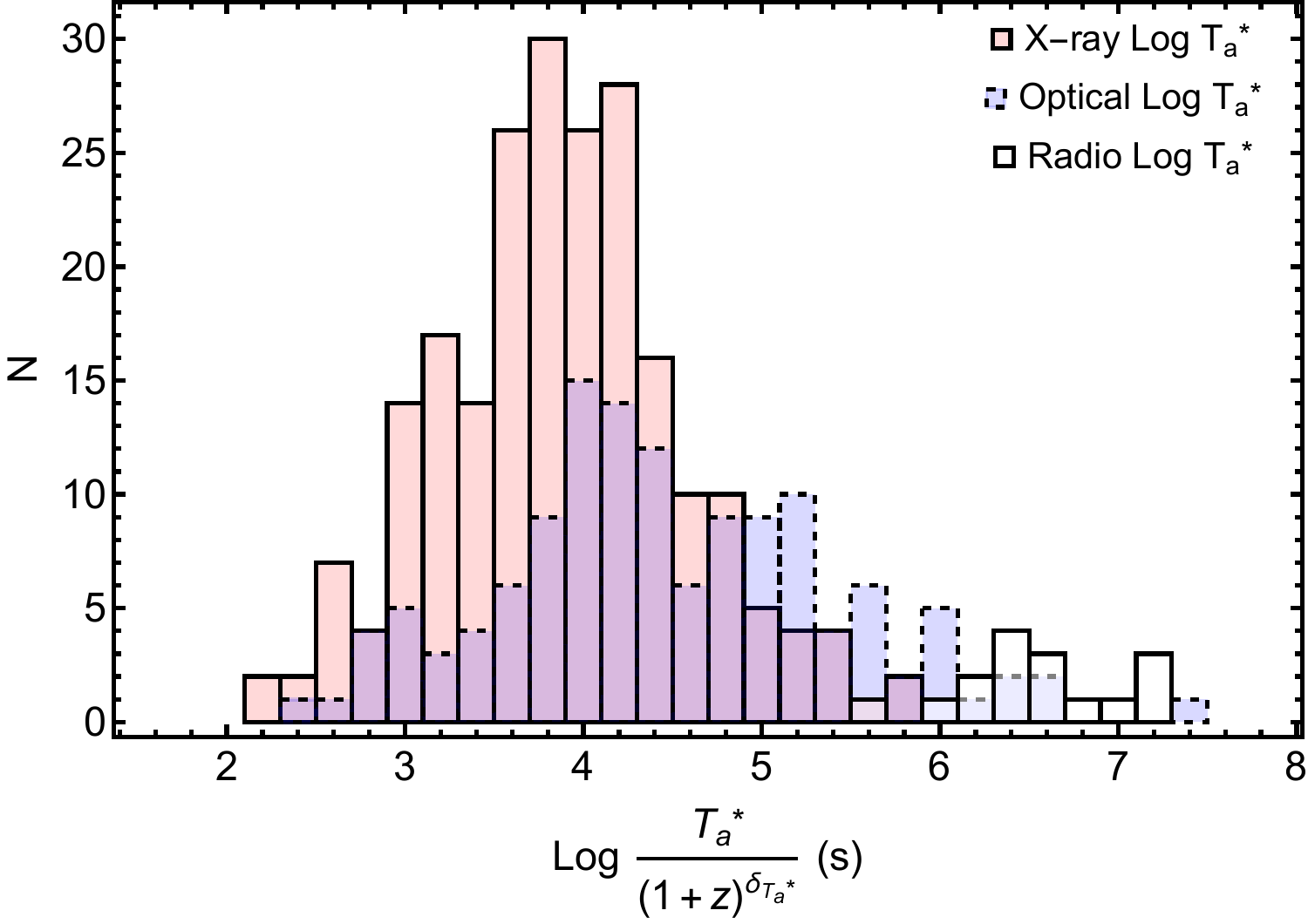}{0.5\textwidth}{}
\fig{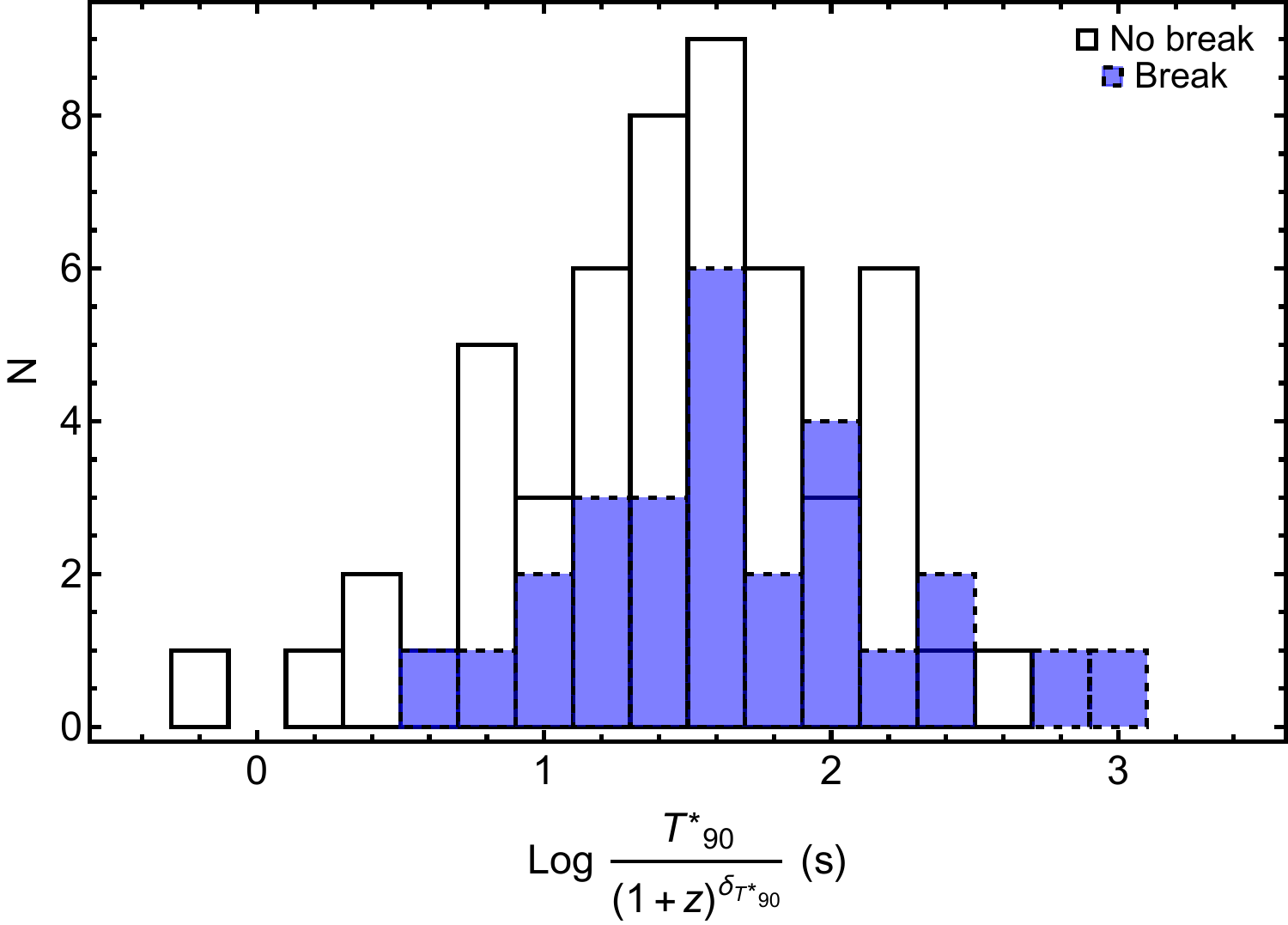}{0.5\textwidth}{}}
\caption{Left panels: Distribution of break time $T_{\rm a}^*$ for X-ray, optical, and radio samples. Right panels: $T^{*}_{90}$ distribution for the sample of 80 GRBs, differentiated between GRBs with or without a break. Lower panels show the same distributions corrected for selection bias.
\label{fig:Tdist}}
\end{figure*}

We compare the $T_a^*$ distribution between the uncorrected and the corrected X-ray, optical, and radio samples to determine if the distributions are drawn from the same parent population (figure~\ref{fig:Tdist}), similar to the work presented in \citet{2020ApJ...905L..26D}. The X-ray sample of 222 GRBs presented in \citet{2020ApJ...904...97D, 2020ApJ...903...18S, 2021PASJ...73..970D} and the extended optical sample (131 GRBs vs 102 GRBs presented in \citet{2020ApJ...905L..26D}) show significant overlap, and a KS test between the two samples without correction gives a value of 0.16 with $p \approx 0.03$, while the comparison of the corrected samples gives a value of 0.31 with $p \approx 0$, indicating a slightly greater difference after correction. However, though both the X-ray and optical samples have an average $T_a^*$ of $\sim 10^4$ seconds, the radio sample has a later average $T_a^*$ of $\sim 10^6$ seconds. A KS test of the radio sample without correction vs. the X-ray and optical samples without correction produces a value of $KS=0.96$ and $KS=0.91$ with a p-value of $\approx 0$, respectively, while a comparison of the corrected samples produces a value of $KS=0.99$ and $KS=0.89$, respectively, with a p-value of $\approx 0$ in both cases, indicating that in both the uncorrected and corrected samples the radio sample is significantly different. We discuss the possible physical mechanisms for this difference in section \ref{sec:discussion}. 

\centerwidetable
\begin{deluxetable*}{lCCCCCCCCCc}[h!]
\tablenum{1}
\tablecaption{Sample of 18 GRBs that display a break and a plateau. Full sample of 82 GRBs used for fitting available online.\label{tab:table1}}
\tablewidth{0.5pt}
\tabletypesize{\scriptsize}
\tablehead{
\colhead{GRB} & \colhead{z} & \colhead{$T_{90}$} & \colhead{log $F_a$} & \colhead{log $T_a$} & \colhead{$\alpha_1$} & \colhead{$\alpha_2$} & \colhead{log $L_a$} & \colhead{$\beta$} & \colhead{ log $E_{\rm iso}$} & \colhead{ref}\\
\colhead{} & \colhead{} & \nocolhead{(s)} & \colhead{(erg/s $cm^2$)} & \colhead{(s)} & \colhead{} & \colhead{} &  \colhead{(erg/s)} & \colhead{} & \colhead{(erg)} & \colhead{}}
\startdata
980329	&	3.9	&	58	&	-16.64	\pm	0.04	&	6.59	\pm	0.09	&	-0.08	\pm	0.06	&	0.83	\pm	0.32	&	43.75	\pm	0.12	&	1.7	\pm	0.17	&	54.32	\pm	0.18	&	[1]	\\
980425	&	0.0085	&	31	&	-15.07	\pm	0.01	&	6.52	\pm	0.01	&	-0.46	\pm	0.05	&	1.55	\pm	0.08	&	38.14	\pm	0.01	&	0.75	\pm	0.17	&	48.20	\pm	0.13	&	[1]	\\
980703	&	0.966	&	90	&	-15.99	\pm	0.11	&	5.95	\pm	0.13	&	-0.33	\pm	0.40	&	0.86	\pm	0.03	&	41.92	\pm	0.12	&	0.46	\pm	0.2	&	52.84	\pm	0.07	&	[1]	\\
000926	&	2.039	&	25	&	-16.78	\pm	0.05	&	6.01	\pm	0.05	&	-0.22	\pm	0.20	&	0.45	\pm	0.15	&	42.24	\pm	0.10	&	0.902	\pm	0.17	&	53.43	\pm	0.11	&	[1]	\\
010222	&	1.477	&	170	&	-17.38	\pm	0.33	&	6.80	\pm	0.28	&	0.06	\pm	0.19	&	1.33	\pm	0.64	&	41.09	\pm	0.34	&	0.902	\pm	0.17	&	54.12	\pm	0.13	&	[1]	\\
011030	&	3	&	1500	&	-16.89	\pm	0.05	&	6.32	\pm	0.07	&	-0.21	\pm	0.20	&	0.91	\pm	0.25	&	42.67	\pm	0.11	&	0.902	\pm	0.17	&	51.69	\pm	0.50	&	[1]	\\
020903	&	0.25	&	13	&	-16.79	\pm	0.29	&	7.16	\pm	0.10	&	0.45	\pm	0.38	&	7.29	\pm	11.20	&	39.53	\pm	0.29	&	0.902	\pm	0.17	&	49.36	\pm	0.12	&	[1]	\\
021004	&	2.33	&	50	&	-16.27	\pm	0.04	&	6.44	\pm	0.04	&	-0.12	\pm	0.08	&	1.30	\pm	0.14	&	42.89	\pm	0.10	&	0.9	\pm	0.17	&	52.58	\pm	0.14	&	[1]	\\
030329	&	0.168	&	63	&	-13.65	\pm	0.02	&	5.94	\pm	0.02	&	-0.09	\pm	0.05	&	1.84	\pm	0.16	&	42.21	\pm	0.02	&	-0.54	\pm	0.02	&	52.26	\pm	0.13	&	[1]	\\
050713B	&	0.55	&	54.2	&	-16.68	\pm	0.06	&	6.31	\pm	0.07	&	0.25	\pm	0.14	&	1.90	\pm	1.05	&	40.53	\pm	0.07	&	0.902	\pm	0.17	&	49.71	\pm	0.41	&	[1]	\\
070612A	&	0.617	&	368.8	&	-16.07	\pm	0.03	&	6.96	\pm	0.05	&	-0.21	\pm	0.07	&	1.52	\pm	0.45	&	41.37	\pm	0.05	&	0.902	\pm	0.17	&	51.96	\pm	0.13	&	[1]	\\
071003	&	1.1	&	150	&	-16.40	\pm	0.12	&	5.82	\pm	0.20	&	-0.12	\pm	0.31	&	0.68	\pm	0.25	&	41.74	\pm	0.14	&	0.902	\pm	0.17	&	51.55	\pm	0.03	&	[1]	\\
111005A	&	0.0133	&	26	&	-15.53	\pm	0.05	&	6.44	\pm	0.04	&	0.00	\pm	0.09	&	2.26	\pm	0.90	&	38.08	\pm	0.05	&	2.03	\pm	0.27	&	48.00	\pm	0.17	&	[2]	\\
111215A	&	2.06	&	796	&	-15.40	\pm	0.02	&	6.06	\pm	0.02	&	0.15	\pm	0.03	&	1.08	\pm	0.08	&	43.45	\pm	0.08	&	0.902	\pm	0.17	&	53.29	\pm	0.12	&	[3]	\\
120326A	&	1.798	&	69.6	&	-16.40	\pm	0.17	&	6.59	\pm	0.27	&	-0.12	\pm	0.30	&	2.04	\pm	2.92	&	42.40	\pm	0.19	&	0.902	\pm	0.17	&	52.51	\pm	0.04	&	[4]	\\
140304A	&	5.283	&	31.2	&	-16.56	\pm	0.04	&	5.73	\pm	0.07	&	-0.11	\pm	0.11	&	0.89	\pm	0.16	&	43.88	\pm	0.32	&	1.1	\pm	0.4	&	53.14	\pm	0.03	&	[5]	\\
141121A	&	1.47	&	33.2	&	-16.72	\pm	0.08	&	6.29	\pm	0.11	&	0.28	\pm	0.18	&	1.10	\pm	0.71	&	41.23	\pm	0.08	&	-0.18	\pm	0.07	&	50.83	\pm	0.16	&	[6]	\\
171010A	&	0.3285	&	104	&	-15.40	\pm	0.02	&	5.52	\pm	0.03	&	-0.11	\pm	0.09	&	1.11	\pm	0.05	&	41.40	\pm	0.02	&	1.9	\pm	0.05	&	53.26	\pm	0.01	&	[7]	\\
\enddata
\tablecomments{Note - Includes GRB ID, redshift, duration, best-fit parameters $F_a$, $T_a^*$, $\alpha_1$, $\alpha_2$, luminosity at time of break $L_a$, radio spectral index $\beta$, and $E_{iso}$. Redshift and $T_{90}$ taken from Greiner (2021) and GCNs when not given in the literature. Best-fit parameters computed from BPL and reported as given by the fit, with the uncertainties at 1 $\sigma$}. $L_a$ computed from best-fit parameters. $\beta$ taken from the literature or set as the average. $E_{iso}$ taken from the literature. References are as follows: 
[1] \citet{2012ApJ...746..156C}, [2] Michalowski et al. (2018), [3] \citet{2013ApJ...767..161Z}, [4] \citet{2015ApJ...814....1L}, [5] \citet{2019ApJ...884..121L}, [6] \citet{2015ApJ...812..122C}, [7] \citet{2019MNRAS.486.2721B}
\end{deluxetable*}

\section{Discussion and conclusions} \label{sec:discussion}

Using the largest compilation of published GRB radio afterglows to date, we achieve two main findings : 1) 18 GRBs resemble a plateau feature in their LCs, and, 2) using this sample, the Dainotti correlation still holds for this radio sample, comparable within 2.1 $\sigma$. After correction for evolutionary effects and removal of outliers, we find that the Dainotti radio correlation is compatible with the corresponding X-ray and optical correlations within 1.5 $\sigma$, with a slope of $a_{\rm rad} = -0.45 \pm 0.47$.
As the slope of the corrected radio correlation agrees with -1 within 1.17 $\sigma$, similar to the slope found in X-ray, this could indicate that the energy reservoir is conserved. One likely candidate for the production of the plateau is a black hole central engine, where the energy injection is driven by fall-back accretion onto the black hole. \citet{2008Sci...321..376K} suggests that the prompt emission is driven by accretion of the outer stellar core, while the plateau phase seen in X-ray LCs is caused by the accretion of the stellar envelope of the progenitor, or is possibly driven by the magnetar. 

Another important finding is that the radio break times occur significantly later than the other wavelengths. On average, the X-ray and optical plateaus last $\sim 10^4$ seconds, while radio plateaus last $\sim 10^6$ seconds. However, there are cases in which longer X-ray plateaus have been seen to last upwards of $\sim 10^5$ seconds, such as GRB 060218 and GRB 980425, detailed in \citet{2017A&A...600A..98D}. The late break times in radio could be a result of the peak of the spectrum appearing in the radio band at later times as the jet decelerates, consistent with the afterglow standard synchrotron shock model. 

On the other hand, the late break could indicate that $T_a^*$ is not the end of a plateau, but a break observed in radio wavelengths. The dynamics of the afterglow emission for an outflow propagating into surrounding constant-density medium has been widely explored \citep{1995ApJ...455L.143S, 1996ApJ...473..204S, 1997ApJ...489L..37S, 1999ApJ...519L..17S}. The outflow transfers a large amount of its energy to this surrounding medium during the deceleration phase. Since the outflow launched into a cone of opening angle $\theta_{\rm j}$ sweeps the surrounding medium, it decelerates, thus increasing the angular size of the emitting region ($\propto 1/\Gamma_{\rm j}$). Once $1/\Gamma_{\rm j}\approx \theta_{\rm j}$, there will be no additional radiating elements from the jet if we assume a ``top hat'' jet, or another structure with a steep drop-off in energy per unit angle outside the core, producing a steepening in the LC. This  ``break" is expected to last from several hours to days \citep[e.g., see][]{2015PhR...561....1K}. After this phase, the energy flux at radio bands is expected to evolve as $F_\nu\propto t^{-p}$ for $\nu_{\rm m} < \nu < \nu_{\rm c}$  and $\propto t^{-\frac{1}{3}}$ for  $\nu < \nu_{\rm m} <  \nu_{\rm c}$ \citep{1999ApJ...519L..17S}. 
In our sample, the $\alpha_2$ values, referring to an evolving slope of the decay phase, range from -0.27 to -0.9 with an average of -0.5, which could potentially support a decay within the $\nu < \nu_{\rm m} <  \nu_{\rm c}$ regime.

Regarding whether the presence of the break is related to the prompt emission, we investigate the distribution between $E_{\rm iso}$ and $T^{*}_{90}$ with and without correction for redshift evolution. We do not observe any trend or clustering in the sample of GRBs that present a break and those without, suggesting that $E_{\rm iso}$ and $T^{*}_{90}$ are not indicators of a jet break within a radio LC. More analysis is needed to determine if there is another physical reason for this distinction.

In conclusion, we find: 
\begin{enumerate}
    \item After correction, the slope of the Dainotti correlation in radio agrees with the X-ray and optical within 1.5 $\sigma$, instead of within 2.1 $\sigma$ when evolutionary effects are not considered. This emphasizes the importance of correcting for selection bias and that the two correlations can be interpreted within the same mechanism if we consider the slope of the correlation as a discriminant among models.
    
    \item The time of break in the radio sample occurs later than in X-ray and optical, and the radio sample is found to be statistically different from the X-ray and optical samples. The late break time can be a result of the passage of the synchrotron characteristic break ($\nu_{m}\propto t^{-\frac32}$) through radio wavelengths 
    during the lateral expansion phase. After the break, the flux at radio bands is expected to evolve first as $F_\nu\propto t^{-\frac{1}{3}}$ for  $\nu < \nu_{\rm m} <  \nu_{\rm c}$ and later, as   $\propto t^{-p}$ for $\nu_{\rm m} < \nu < \nu_{\rm c}$ \citep{1999ApJ...519L..17S}. Another plausible explanation could be that the flux that dominates the radio bands is emitted in a wider decelerated shell \citep[e.g., see][]{2014MNRAS.437.1821M, 2013ApJ...778..107S, 2021ApJ...907...78F}.
    \item Analysis of the $E_{\rm iso}$ and $T^*_{90}$ of 80 GRBs demonstrates that GRBs with and without a break in the LC appear to be drawn from the same parent population.
    \end{enumerate}

\section{Acknowledgements}
\begin{acknowledgements}
This work was made possible in part by the United States Department of Energy, Office of Science, Office of Workforce Development for Teachers and Scientists (WDTS) under the Science Undergraduate Laboratory Internships (SULI) program. We thank Dr. Cuellar for managing the SULI program at Stanford National Accelerator Laboratory. We also acknowledge the National Astronomical Observatory of Japan for their support in making this research possible, as well as Aleksander Lenart for helpful discussions during the writing of this paper. PC acknowledges support of the Department of Atomic Energy, Government of India, under project no. 12-R\&D-TFR-1155 5.02-0700. 
\end{acknowledgements}

\bibliography{radioref}{}
\bibliographystyle{aasjournal}

\end{document}